\documentclass[12pt]{article}
\usepackage{amssymb}
\usepackage{amsfonts}
\usepackage{latexsym}
 
\newcommand{\be}{\begin{equation}}
\newcommand{\bdm}{\begin{displaymath}}
\newcommand{\bea}{\begin{eqnarray}}
\newcommand{\ee}{\end{equation}}  
\newcommand{\edm}{\end{displaymath}}
\newcommand{\eea}{\end{eqnarray}}
\newcommand{\nn}{\nonumber}
\newcommand{\p}{\partial} 

\begin{document}

\begin{flushright}
NYU-TH/04-03-07\\
hep-th/0403197
\end{flushright}

\vspace{0.5cm}
\begin{center}
\boldmath{\Large \bf $D$-Brane Interactions in a 
Gravitational Shock Wave Background}\unboldmath\\
\vspace{1cm}
{\large 
L.~Girardello$^{a,}$\footnote{e-mail: luciano.girardello@mib.infn.it}, 
C.~Piccioni$^{b,}$\footnote{e-mail: carlo.piccioni@physics.nyu.edu}, 
M.~Porrati$^{b,}$\footnote{e-mail: massimo.porrati@nyu.edu}} 

\vspace{0.5cm}
$^a${\it Universit\`{a} di Milano-Bicocca and INFN\\Piazza Della Scienza 3, I-20126 Milano, Italy}\\
\vspace{0.3cm}
$^b${\it Department of Physics, New York University\\
4 Washington Place, New York NY 10003, USA} \\
\end{center}
\bigskip

\begin{center}
\bf Abstract 
\end{center}
{ 
We study $D-$branes in the background of a gravitational shock wave. 
We consider the case of parallel $D-$branes located on opposite sides with 
respect to the shock wave. Their interaction is studied by evaluating  the cylinder diagram using the boundary states technique. Boundary states are defined at each $D-$brane and their scalar product is evaluated after propagation through the shock wave. Taking the limit where the gravitational shock wave vanishes we show that the amplitude evaluated is consistent with the flat space$-$time result.}

\newpage  
  
\section{Introduction}

The description of time dependent space-times in string theory is important 
for many reasons (applications to cosmology, the understanding of singularities etc.). An example of time dependent background  is made in  ~\cite{c1} where time dependent $D$-branes, carrying electromagnetic pulses, are considered in Minkowski space-time.
\\
In this paper, we study an interesting and simple example of time dependent 
background; namely, we introduce flat $D$-branes in the background of a 
gravitational shock wave. This system is exactly soluble, since away from the 
shock wave the space-time is flat.
\\
The propagation of strings in this background, as described in ~\cite{c2,c3}, 
is studied in the light-cone gauge  where the equations of motion can be 
exactly solved. Strings freely propagate until they intersect with the shock 
wave. This allows for the use of the Minkowski formulation in terms of 
oscillator expansions, which undergo an $S$ matrix transformation when the 
interaction takes place. An updated and complete reference for strings in shock and plane wave backgrounds can be found in ~\cite{c4}.
\\
As explained in ~\cite{c2} the fermionic degrees of freedom are not modified by the shock wave background. This is true as long as the shock wave background is purely bosonic as in the case here considered. In a background with fermionic components, as in the supergravity shock wave backgrounds, interesting interactions emerge, for instance transfer of fermionic degrees of freedom between the string and the background. This was found and derived in ~\cite{c5}. These consideration are beyond the purpose of this letter, in the following we consider the bosonic shock wave background and therefore restrict ourselves to the bosonic string variables. 
\\
Away from the shock wave, the D-branes are described by standard 
flat hyperplanes. We shall use the formalism developed in ~\cite{c6} to 
describe strings attached to them.
\\
Next we define boundary states for our D-brane configuration. They are easily 
obtained from the flat space-time formulation, where they implement 
closed-string boundary conditions at fixed world-sheet time. From this fact 
and the light-cone gauge choice, it follows that the light-cone coordinate 
parametrizing the world-sheet time  satisfies Dirichlet boundary conditions.
\\
This defines a Euclidean world-volume, so that $Dp$-branes can be seen as 
$D(p+1)$-instantons, i.e. objects living at fixed world 
sheet-time ~\cite{c6,c7}.
\\
In this formalism the shock wave is also at fixed world-sheet time, so the 
configurations considered here are $D(p+1)$-instantons parallel to it.
The most interesting configuration is that where the $D(p+1)$-instantons
are located on opposite sides with respect to the shock wave.
\\
The paper is structured as follows, in Section $2$, we recall preexisting 
results on the propagation of closed strings in the shock wave background. 
In Section $3$, we introduce $D(p+1)-$instantons, boundary states, and 
the string propagator ~\footnote{Note that the propagation implies crossing 
the shock wave, resulting in a time dependent theory.}. We then 
evaluate the tree level amplitude for the exchange of a  closed string in 
terms of an overlap between two boundary states on opposite sides of the 
shock wave. In Section $4$, we discuss the  open string approach, and discuss
the interpretation of the result of Section $3$ in terms of the one-loop open 
string amplitude. This calculation, not performed here, is the crucial 
consistency test for our system.

\section{Strings in the shock wave geometry}

The background of a gravitational shock wave is
\be
ds^2=g_{\mu \nu}dx^{\mu}dx^{\nu}=-2dudv+(d\vec x)^2+f(\vec x)\delta(u)du^2\, , 
\ee
where $\vec x =(x^{1}...x^{d-2})$ and $u=(x^{0}-x^{d-1})/\sqrt{2}$, $v=(x^{0}+x^{d-1})/ \sqrt{2}$ are light cone coordinates.
\\
In order to obtain vacuum solutions to the Einstein equations the profile $f(\vec x)$ must be harmonic in the transverse coordinates, $\Delta_{T}f=0$. 
In fact, the only non vanishing component of the Ricci tensor is $R_{uu}=-\frac{1}{2}\Delta_{T}f\delta(u)$.
\\
We will use the simple quadratic profile  
\be
f(\vec x)=\sum_{I=1}^{d-2} a^{I}(x^{I})^2\, ,
\ee
with the constraint
\be
\sum_{I=1}^{d-2}a^{I}=0\, .
\ee
\\
The propagation of a classical  string in this background, ~\cite{c2,c3}, is  described by the Lagrangian density
\be
{\cal L}=-\frac{1}{2\pi}g_{\mu \nu}\p_{\alpha}X^{\mu}\p^{\alpha}X^{\nu}\, ,
\ee
which gives the free equation of motion for the light-cone embedding $U$
\be
\left(-\p_{\tau}^2+\p_{\sigma}^2\right)U(\tau,\sigma)=0\, .
\ee
This allows the light-cone gauge choice $U=p^{u}\tau +u$ .
\\
The equations of motion for the other coordinates then simplify to the form:
\be
\left(-\p_{\tau}^2+\p_{\sigma}^2\right)X^{I}(\tau,\sigma)=-\frac{1}{2}p^{u}\p_{I}f(\vec X)\delta(\tau+\frac{u}{p^{u}})\, ,
\ee
\be
\left(-\p_{\tau}^2+\p_{\sigma}^2\right)V(\tau,\sigma)=-\frac{1}{2}f(\vec X)\delta^{\prime}(\tau+\frac{u}{p^{u}})-\p_{I}f(\vec X)\p_{\tau}X^{I}\delta(\tau+\frac{u}{p^{u}})\, .
\ee
Let us consider a closed string. For $\tau >-\frac{u}{p^{u}}$ (respectively $\tau <-\frac{u}{p^{u}}$), we obtain the usual flat space solutions, either
$ X^{I}(\tau,\sigma)=X^{I}_{>}(\tau,\sigma)$ or 
$ X^{I}(\tau,\sigma)=X^{I}_{<}(\tau,\sigma)$, with 
\be
X^{I}_{\gtrless}(\tau , \sigma )= x^{I}_{\gtrless}+p^{I}_{\gtrless}\tau+\frac{i}{2}\sum_{n\neq 0}\frac{1}{n} \left( e^{-2in(\tau - \sigma )}\alpha^{I}_{n \gtrless}+e^{-2in(\tau+\sigma)} \tilde \alpha^{I}_{n \gtrless} \right)\, ;
\ee
an analogous relation holds for the coordinate $V$.\\
Solving Eq.~$(6)$ for each Fourier mode and imposing the continuity of $\vec X$ at $\tau =-\frac{u}{p^{u}}$ one finds the following relations for the expansion modes  
\bea
x^{I}_{>}&=&x^{I}_{<}+\frac{u}{2\pi}\int_{0}^{\pi}d\sigma \p_{I}f\left(\vec X(-\frac{u}{p^{u}},\sigma)\right)\, ,
\\
p^{I}_{>}&=&p^{I}_{<}+\frac{p^u}{2 \pi}\int_{0}^{\pi}d \sigma \p_{I}f\left(\vec X(-\frac{u}{p^{u}},\sigma)\right)\, ,
\\
\alpha^{I}_{n >}&=&\alpha^{I}_{n <}+\frac{p^{u}}{4\pi}e^{-2in\frac{u}{p^{u}}}\int_{0}^{\pi}d\sigma e^{-2in\sigma }\p_{I}f\left(\vec X(-\frac{u}{p^{u}},\sigma )\right)\, ,
\\
\tilde \alpha^{I}_{n >}&=&\tilde \alpha^{I}_{n <}+\frac{p^{u}}{4\pi}e^{2in\frac{u}{p^{u}}}\int_{0}^{\pi}d\sigma e^{2in\sigma }\p_{I}f\left(\vec X(-\frac{u}{p^{u}},\sigma )\right)\, .
\eea
For the $V$ coordinate, we find that Eq.~$(7)$ can be derived imposing Eq.~$(6)$ on the Virasoro constraints, which read
\bea
2p^{u}\p_{\tau}V&=&(\p_{\tau}X^{I})^2+(\p_{\sigma}X^{I})^2+p^{u}\delta(\tau+\frac{u}{p^{u}})\, ,
\\
p^{u}\p_{\sigma}V&=&\p_{\tau}X^{I}\p_{\sigma}X^{I}\, .
\eea
\\
This implies, for either $\tau>-\frac{u}{p^{u}}$ or $\tau<-\frac{u}{p^{u}}$,
the usual light cone gauge relations for the modes of $V$ 
\be
\alpha ^{v}_{n\gtrless}=\frac{1}{p^{u}}\sum_{m=-\infty}^{+\infty}\alpha^{I}_{n-m \gtrless}\alpha^{I}_{m \gtrless}\, ,
\ee
where $2\alpha ^{v}_{0\gtrless}=p^{v}_{\gtrless}$. An analogous relations 
holds for $\tilde \alpha ^{v}_{n\gtrless}$.
\\
Note that while Eq.~$(7)$ is ambiguous because of the discontinuity of 
$\p_{\tau}X^{I}$ at $\tau=-\frac{u}{p^{u}}$, Eqs.~$(13)$ and $(14)$ are not, so they should be used to determine $V$ instead of Eq.~$(7)$.
\\
By direct integration of Eq.~$(14)$, and by solving for the zero modes in  Eq.~$(15)$, 
we find the relation 
\be
v_{>}=v_{<}+\frac{1}{2\pi}\int_{0}^{\pi}d\sigma f\left(\vec X_{<}(-\frac{u}{p^{u}},\sigma)\right)\, ,
\ee
where we have chosen 
\be
\p_{\tau}X^{I}(-\frac{u}{p^{u}},\sigma )=-\frac{p^{u}}{4}\p_{I}f\left(\vec X(-\frac{u}{p^{u}},\sigma )\right).
\ee
With this choice we also have $p^{v}_{>}=p^{v}_{<}$.
\\
The quantization of the theory is done by interpreting the expansion modes as 
operators satisfying canonical commutation relations. The ``in'' and ``out'' 
modes should be related by a unitary transformation, the  $S$ matrix.
\\
It is easy to check that its explicit form is:  
\be
S=\exp\left[ \frac{-ip^{u}}{2\pi}\int_{0}^{\pi}d\sigma 
f\left(\vec X(-\frac{u}{p^{u}} ,\sigma )\right)\right] .
\ee
It gives the transformation relations
\be
\alpha_{>}=S\alpha_{<} S^{\dagger}\, ,
\ee
where  $\alpha$ denotes any expansion mode. This formula holds also before 
making the choice in Eq.~$(17)$.
\\
In the case of the profile in Eq.~$(2)$ the $S$ matrix takes the form
\bdm 
S=\exp\Bigg\{-\frac{i}{2}a^{I}p^{u}\left[(x^{I}_{<}-\frac{u}{p^{u}}p^{I}_{<})^2\right.
\edm
\be
\left.  
+ \frac{1}{2}\sum_{n=1}^{\infty}\frac{1}{n}\left(e^{2in\frac{u}{p^{u}}}a^{I}_{n<}-e^{-2in\frac{u}{p^{u}}}\tilde a^{I\dagger}_{n<}\right)\left(e^{-2in\frac{u}{p^{u}}}a^{I\dagger}_{n<}-e^{2in\frac{u}{p^{u}}}\tilde a^{I}_{n<}\right)
\right]\Bigg\}\, ,
\ee
where in  Eq.~$(8)$ we have chosen the $<$ expansion modes. 
Here and in the following we defined $\sqrt{n}a^{I}_{n}=\alpha ^{I}_{n}$. 

\section{ $D$-branes in  the shock wave geometry}

The picture so far is of a closed string interacting with the shock wave only at $u=0$. The string is in flat space-time everywhere else. 
\\
Applying  the formalism introduced in \cite{c6} it is then possible to choose Neumann or Dirichlet boundary conditions for the closed string  at some fixed world-sheet time away from the shock wave.
\\
We note that, as in the flat space-time case, at fixed world-sheet time the $U$ coordinate is Dirichlet by definition and the $V$ coordinate is Dirichlet 
too, because of the Virasoro constraint of Eq.~$(14)$.
\\
The boundary conditions for the other coordinates can be chosen either Neumann or Dirichlet, say Neumann for $\alpha=1...p+1$ and Dirichlet for $i=p+2...d-2$. 
\\
This defines a Euclidean $Dp$-brane or $D(p+1)-$instanton located at some 
$u=u_{0}\ne 0$ ,$ v=v_{0}$ , $ x{^i}=y^{i}$. It is then straightforward to 
define the boundary states as in the flat space-time case \cite{c8}
\bea
\p_{\tau} X^{\alpha}|_{\tau=0}|B\rangle&=&0\, ,   
\\
X^{i}|_{\tau=0}|B\rangle&=& y^{i}|B\rangle\, ,           
\\
U|_{\tau=0}|B\rangle&=& u_{0}|B\rangle\, , 
\\
V|_{\tau=0}|B\rangle&=& v_{0}|B\rangle\, .
\eea
These conditions are implemented by the well known boundary state
\bdm
|B;u_{0},v{_0},y^{i}\rangle=|u=u_{0}\rangle|v=v{_0}\rangle|x^{i}=y^{i}\rangle|p^{\alpha}=0\rangle
\edm
\be
\times \exp\left[\sum_{n=1}^{\infty} \epsilon _{p}^{I}a^{I \dagger}_{n}\tilde a^{I \dagger}_{n} \right]|0\rangle|\tilde 0\rangle\, ,
\ee
where $\epsilon _{p}^{I}=-1$  for $ I=1...p+1$ and $1$ elsewhere.
\\
Suppose now we want to evaluate the interaction between two $D(p+1)-$instantons located on opposite sides of the shock wave.
\\
Then, one will be located at $u=u_{0<}<0$, $v=v_{0<}$, $x^{i}=y^{i}_{<}$, and described by the boundary state 
\bdm
|B_{<};u_{0<},v_{0<},y^{i}_{<}\rangle=|u=u_{0<}\rangle|v_{<}=v_{0<}\rangle|x^{i}_{<}=y^{i}_{<}\rangle|p^{\alpha}_{<}=0\rangle
\edm
\be
\times \exp\left[\sum_{n=1}^{\infty} \epsilon _{p}^{I}a^{I \dag}_{n<}\tilde a^{I \dag}_{n<} \right]|0\rangle_{<}|\tilde 0\rangle_{<}\, .
\ee
The other one will be located at $u=u_{0>}>0$, $v=v_{0>}$, $x^{i}=y^{i}_{>}$, and described by the boundary state
\bdm
|B_{>};u_{0>},v_{0>},y^{i}_{>}\rangle=|u=u_{0>}\rangle|v_{>}=v_{0>}\rangle|x^{i}_{>}=y^{i}_{>}\rangle|p^{\alpha}_{>}=0\rangle
\edm
\be
\times \exp\left[\sum_{n=1}^{\infty} \epsilon _{p}^{I}a^{I \dag}_{n>}\tilde a^{I \dag}_{n>} \right]|0\rangle_{>}|\tilde 0\rangle_{>}\, .
\ee
Then the  interaction amplitude, described by the cylinder amplitude, is an overlap between the boundary states and  takes the form
\be
{\cal A}(\Delta u,\Delta v,y^{i}_{<},y^{i}_{>})=\langle B_{>};u_{0>},v_{0>},y^{i}_{>}|D|B_{<};u_{0<},v_{0<},y^{i}_{<}\rangle\, ,
\ee
where $\Delta u,\Delta v$ are the separations in the light-cone directions  and $D$ denotes the propagator.
\\
To evaluate the propagator, we first note that the Hamiltonian density of the theory is explicitly $\tau$ dependent
\be
{\cal H}={\cal H}_{0}+\frac{p^{u}}{2\pi}f\left(\vec X(-\frac{u}{p^{u}},\sigma)\right)\delta(\tau+\frac{u}{p^{u}})\, ,
\ee
where 
\be
{\cal H}_{0}=\frac{1}{2\pi}\left[-2p^{u}\p_{\tau}V+(\p_{\tau}X^{I})^2+(\p_{\sigma}X^{I})^2\right]
\ee
is the flat space-time Hamiltonian density.
\\
Then the Hamiltonian takes the form
\be
H=H_{0}+\frac{p^{u}}{2\pi}\int_{0}^{\pi}d\sigma f\left(\vec X(-\frac{u}{p^{u}},\sigma)\right)\delta(\tau+\frac{u}{p^{u}})\, ,
\ee
and away from the shock wave $H_{0}$ is explicitly given  by
\be
H_{0}=H_{\gtrless}=-p^{u}p^{v}_{\gtrless}+ \frac{(p^{I}_{\gtrless})^2}{2}+2\sum_{n=1}^{\infty}n(a^{I \dag}_{n \gtrless}a^{I}_{n \gtrless}+ \tilde a^{I \dag}_{n \gtrless}\tilde a^{I}_{n \gtrless})\, . 
\ee
As we expect, the string is propagating in Minkowski space-time away from the shock wave. 
\\
To find the correct form of the propagator we note that the Hamiltonian 
$H=H_{0}+V\delta(t-t_{0})$ contains an instantaneous interaction term
$V\delta(t-t_{0})$, while $H_{ 0}$ is time independent. By solving the 
Schr\"oedinger equation for the propagator $G(t)$: $i\dot{G}\!=\!HG$, 
one finds $G(t)=\theta(t_{0}-t)e^{-iH_{0}t}\!+\theta(t-t_{0})e^{-iV}
e^{-iH_{0}t}$.
\\
We can then formally  write the  string propagator as 
\be
D=\int_{0}^{-\frac{u}{p^{u}}}d\tau e^{-i\tau H_{0}}+e^{\frac{-ip^{u}}{2\pi}\int_{0}^{\pi}d\sigma f\left(\vec X(-\frac{u}{p^{u}},\sigma)\right)}\int_{-\frac{u}{p^{u}}}^{\infty}d\tau e^{-i\tau H_{0}}\, .
\ee
We immediately recognize the $S$ matrix
\be
S=\exp\left[\frac{-ip^{u}}{2\pi}\int_{0}^{\pi}d\sigma f\left(\vec X(-\frac{u}{p^{u}},\sigma)\right)\right]\, .
\ee
This is no surprise since propagating past the shock wave we need a change of 
basis to describe the system. At this point, we no longer need to transform the oscillators, so from now on we will drop the $\gtrless$ subscript.
\\
Since we want to propagate $|B_{<};u_{0<},v_{0<},y^{i}_{<}\rangle$ until some
time $\tau$ after the interaction with the shock wave, we need only the second term of the propagator.
\\
Shifting $\tau $ and performing a Wick rotation, we get,  using standard 
boundary-state techniques ~\cite{c8}, and denoting the Euclidean time again 
with  $\tau$
\bdm 
D|B_{<};u_{0<},v_{0<},y^{i}_{<}\rangle= 
\edm
\bdm
S\int_{0}^{\infty}d\tau \int_{-\infty}^{+\infty}\frac{dp^{u}dp^{v}}{(2\pi)^2}e^{(\tau-i\frac{u_{0<}}{p^{u}})p^{u}p^{v}+i(p^{v}u_{0<}+p^{u}v_{0<})}|p^{u}\rangle|p^{v}\rangle
\edm
\bdm 
\times e^{-(\tau-i\frac{u_{0<}}{p^{u}})\frac{(p^{i})^2}{2}}|x^{i}=y^{i}_{<}\rangle|p^{\alpha}=0\rangle 
\edm
\be
\times \exp\left[\sum_{n=1}^{\infty} e^{-4(\tau-i\frac{u_{0<}}{p^{u}}) n} \epsilon _{p}^{I}a^{I \dag}_{n}\tilde a^{I \dag}_{n}\right]|0\rangle_{<}|\tilde 0\rangle_{<}\, .
\ee
Then, by acting with $S$ on  $|B_{>};u_{0>},v_{0>},y^{i}_{>}\rangle$, 
integrating over $p^{u}$ and $p^{v}$ and performing  the change of variables 
$t=\frac{\Delta u}{u_{0>}}\tau $, the amplitude of Eq.~$(28)$ becomes
\bdm
{\cal A}(\Delta u,\Delta v,y^{i}_{<},y^{i}_{>})=\int_{0}^{\infty}\frac{dt}{2\pi t}e^{\frac{2\Delta u \Delta v}{t}}\langle p^{\alpha}=0|e^{-\frac{\Delta u a^{\alpha}}{t}(x^{\alpha}+i\frac{u_{0>}t}{2\Delta u}p^{\alpha})^2}|p^{\alpha}=0\rangle \edm
\bdm
\times \langle y^{i}_{>}|e^{-\frac{\Delta u a^{i}}{t}(x^{i}+i\frac{u_{0>}t}{2\Delta u}p^{i})^2}e^{-\frac{t}{4}(p^{i})^2}|y^{i}_{<}\rangle \nn
\edm
\bdm
\times \langle \tilde 0|\langle 0|\exp\left[\sum_{n=1}^{\infty} \epsilon _{p}^{I}a^{I}_{n}\tilde a^{I}_{n} \right] 
\edm
\bdm
\times \exp \left[-\frac{\Delta u a^{I}}{2t}\sum_{n=1}^{\infty}\frac{1}{n}\left(e^{\frac{u_{0>}}{\Delta u}nt}a^{I}_{n}-e^{-\frac{u_{0>}}{\Delta u}nt}\tilde a^{I\dagger}_{n}\right)\left(e^{-\frac{u_{0>}}{\Delta u}nt}a^{I\dagger}_{n}-e^{\frac{u_{0>}}{\Delta u}nt}\tilde a^{I}_{n}\right)\right]
\edm
\be
\times \exp\left[\sum_{n=1}^{\infty} e^{-2nt}\epsilon _{p}^{I}a^{I \dag}_{n}\tilde a^{I \dag}_{n} \right]|0\rangle|\tilde 0\rangle\, .
\ee
To evaluate the first two scalar products in Eq.~$(36)$ we note that the operator  $(x^{I}+i\frac{u_{0>}t}{2\Delta u}p^{I})$ can be normalized to a lowering operator $c$ with $[c,c^{\dagger}]=1$, then using coherent states expansions we get
\be 
\langle p^{\alpha}=0|e^{-\frac{\Delta u a^{\alpha}}{t}(x^{\alpha}+i\frac{u_{0>}t}{2\Delta u}p^{\alpha})^2}|p^{\alpha}=0\rangle= \sqrt{\frac{2\pi}{u_{0>}a^{\alpha}}}
\ee
and 
\bdm
\langle y^{i}_{>}|e^{-\frac{\Delta u a^{i}}{t}(x^{i}+i\frac{u_{0>}t}{2\Delta u}p^{i})^2}e^{-\frac{t}{4}(p^{i})^2}|y^{i}_{<}\rangle=\sqrt{\frac{\Delta u}{\pi t(\Delta u-a^{i}u_{0>}u_{0<})}}
\edm
\be
\times \exp\left\{-\frac{\Delta u\left[(1-a^{i}u_{0<})(y^{i}_{>})^2-2y^{i}_{>}y^{i}_{<}+(1+a^{i}u_{0>})(y^{i}_{<})^2\right]}{t(\Delta u-a^{i}u_{0>}u_{0<})}\right\}\, .
\ee
For the third scalar product we note that for a couple  of independent 
harmonic oscillators $a,\tilde a$ with  $[a,a^{\dagger}]=1$ and  $[\tilde a,\tilde a^{\dagger}]=1$, we have 
\be
\langle \tilde 0|\langle 0|e^{ua\tilde a}e^{-g(ca-\frac{\tilde a^{\dagger}}{c})(\frac{a^{\dagger}}{c}-c\tilde a)}e^{va^{\dagger}\tilde a^{\dagger}}|0\rangle|\tilde 0\rangle=\frac{1}{1-uv+g(1-\frac{u}{c^2})(1-c^2v)}  ,
\ee
where $u,v,g,c$  are some real numbers. 
The coherent-state product over the oscillators $a_{n}^{I}, \tilde{a}_{n}^{I}$, factorizes into a product that reads:
\be
\prod_ {I=1}^{d-2}\prod _{n=1}^{\infty}(1-e^{-2nt})^{-1}\left [1+\frac{\Delta u a^{I}}{2nt}\frac{(1-\epsilon ^{I}e^{-\frac{u_{0>}}{\Delta u}2nt})(1-\epsilon ^{I}e^{\frac{u_{0<}}{\Delta u}2nt})}{1-e^{-2nt}}\right]^{-1}\, .
\ee
For Neumann directions we have $\epsilon ^{\alpha}\!=-1$, with $ \alpha =1...p+1$, so that $(40)$ simplifies to
\be
\prod_ {\alpha=1}^{p+1}\prod _{n=1}^{\infty}(1-e^{-2nt})^{-1}\left[1+\frac{\Delta u a^{\alpha}}{nt}\frac{\cosh(\frac{-u_{0<}}{\Delta u}nt)\cosh(\frac{u_{0>}}{\Delta u}nt)}{\sinh(nt)}\right]^{-1}\, .
\ee
For Dirichlet directions $\epsilon ^{i}\!=1$ with $i=p+2...d-2$ so we have
\be
\prod_ {i=p+2}^{d-2}\prod _{n=1}^{\infty}(1-e^{-2nt})^{-1}\left[1+\frac{\Delta u a^{i}}{nt}\frac{\sinh(\frac{-u_{0<}}{\Delta u}nt)\sinh(\frac{u_{0>}}{\Delta u}nt)}{\sinh(nt)}\right]^{-1}\, .
\ee
We immediately note that plugging Eqs.~$(37),(38),(40)$ into Eq.~$(36)$ and taking the limit $a^{I}=0$ for which the shock wave goes to zero, we get the usual result for strings in Minkowski space time: the amplitude has a Gaussian dependence  on the square of the distance between the two $D(p+1)$-instantons and Eq.~$(40)$ gives the Minkowski partition function.
\\
This represents a first consistency check. To have full proof of the 
correctness of the result we would need to evaluate the same object in the open string channel and get the same answer. 

\section{Conclusions}

We have evaluated the static interaction between two $D(p+1)$-instantons 
located on the opposite sides of a gravitational shock wave and shown that, 
in the flat space-time limit, our formulae reduce to the standard results.
\\
The final  consistency check would come from computing the same interaction in the open-string channel, obtained tansforming the cylinder diagram to an annulus by a modular transformation. This channel is also appropriate for the  study of the singularities of the amplitude as a function of the distance of the branes and consequent physical interpretation of the result. This is not presented in this letter.
\\
In our setup, open strings end on the $D(p+1)$-instantons of Section 3. 
One must then compute the cylinder diagram as a one-loop open string vacuum
diagram. The open strings are defined by introducing a light-cone gauge defined by 
\be 
U=\frac{\Delta u}{\pi}\left(\sigma + \frac{u_{0<}\pi}{\Delta u}\right)\, .
\ee
This choice is allowed by Eq.~$(5)$; it corresponds to interchanging 
$\tau$ and $\sigma$ in the closed string channel. 
The string constructed in this manner intersects the shock wave at 
$\sigma=-\frac{u_{0<}\pi}{\Delta u}$, and lives in Minkowski space 
everywhere else. 
Appropriate boundary conditions can be chosen for the other coordinates at 
$\sigma=0,\pi$, where the string lies on the $D(p+1)$-instantons.
\\
By substituting Eq.~$(43)$ in the Lagrangian density given in Eq.~$(4)$, 
we find the equations of motion for the transverse coordinates. 
It can be shown that regardless the boundary conditions they are equivalent to
an infinite set of classical, coupled harmonic oscillators. Evaluation of  
the cylinder diagram as a trace over open string states thus involves  the 
usual open string partition function with appropriate frequencies. By 
decoupling the oscillators, we find a set of nontrivial equations for these 
frequencies, which make it difficult to evaluate  the amplitude in this 
channel. This is why, up to now, we have not been able to cast the 
open-string computation in a closed form. The closed-string results reported 
here are nevertheless, in the authors' opinion, simple and interesting enough
to warrant being communicated.

\subsection*{Acknowledgments}
M.P. is supported in part by NSF  grants PHY-0245068 and PHY-0070787. L.G. is partially supported by INFN and MURST under contract 2003-023852-008, and by the European Commission TMR program HPRN-CT-2000-00131.

\end{document}